\documentclass[pre,aps,floatfix,superscriptaddress,twocolumn]{revtex4-1}
\usepackage{graphicx}
\usepackage{dcolumn}
\usepackage{latexsym}
\usepackage{hyperref}
\usepackage{amsmath, amsthm, amssymb}
\usepackage{epsfig}
\usepackage{bm}
\usepackage{geometry}
\usepackage{tikz}
\usepackage[dvipsnames]{xcolor}
\begin{document}

\title{{\bf From LRO to Disorder via QLRO in Spatially Inhomogeneous Polar Flock}}

\author{Anish Kumar}
\email[]{anishkumar.rs.phy22@itbhu.ac.in} 
\affiliation{Department of Physics, Indian Institute of Technology(BHU), 
Varanasi- 221005, India.}

\author{Vivek Semwal}
\email[]{viveksemwal.rs.phy17@itbhu.ac.in} 
\affiliation{RIKEN Center for Biosystems Dynamics Research, Kobe 650-0047, Japan.}

\author{Shradha Mishra}
\email[]{smishra.phy@itbhu.ac.in}
\affiliation{Department of Physics, Indian Institute of Technology(BHU), 
Varanasi- 221005, India.}

\begin{abstract}
We study the collective behavior of a polar flock in an inhomogeneous environment in two-dimensions. The inhomogeneity is modelled by introducing regions at random locations on the substrate with higher noise but  accessible for the flock to move. Hence inside such regions the particles orientation get randomised.  Such inhomogeneities  are different from  the physical disorder, which obstructs the space for the incoming particles.  The study focuses on how the phase behavior of polar flock changes by tuning the packing fraction of inhomogeneity. As packing fraction increases, the system crosses over from long-range to quasi-long-range order and ultimately to a disordered phase, while the order–disorder transition for flocking changes from discontinuous to continuous. The resultant phase behavior of polar flock patterns here is comparable to that exhibited in the presence of physical disorder.
\end{abstract}

\maketitle 

\textit{\textbf{Introduction}}: Collective motion is a universal phenomenon across scales, from bacterial colonies to animal herds, where flocks are much larger than individual constituents. These agents, known as self-propelled particles (SPPs), convert internal energy into mechanical work to move autonomously. The minimal Vicsek model \cite{vicsek1995novel} demonstrated that SPPs can exhibit long-range order (LRO) even in two dimensions, apparently violating the Mermin–Wagner theorem \cite{vicsek1995novel,toner1995long}. While most studies focus on homogeneous environments, some works on SPPs in the presence of inhomogeneous environments show a significant impact of inhomogeneity on flocking behavior, like SPPs exhibit quasi-long-range order (QLRO) in the presence of quenched physical disorder \cite{chepizhko2013optimal,das2018polar,toner2018hydrodynamic}.\\

Previous studies have primarily focused on inhomogeneity arising from physical disorder, i.e. restricted volumes or random rotators (e.g., road signs) that attempt to align SPPs in specific directions. However, it can take other forms, including patterned activity \cite{mishra2023active} or variations in temperature. Here, we introduce spatial inhomogeneity by introducing some regions of high noise in the space, where SPPs can move freely, but experience high noise that disrupts local alignment. In contrast, the rest of the space maintains low noise to facilitate flocking. This represents a spatially inhomogeneous environment. Through our study we focus on the  effect of fluctuations arise while the flock on move rather than purely physical obstacles.\\

In this study, we address how spatial inhomogeneity, introduced by circular regions of high noise, affects the ordering of SPPs and the nature of the flocking transition. The SPPs interact via Vicsek alignment, following the motion of their neighbors with a noise experiencing throughout the system, which is maximum in circular regions. Our key findings are: (i) as the packing fraction of inhomogeneity increases, the system transitions from long-range order (LRO) to quasi-long-range order (QLRO) and finally to a disordered state; (ii) the inhomogeneity changes the nature of the flocking transition from discontinuous to continuous. The overall effects of the inhomogeneity are largely parallel to those of physical disorder.

\textit{\textbf{Model}}: We consider a collection of $N$ SPPs moving on a square substrate of side $L$ with a constant speed $v_o$ under periodic boundary conditions. Each SPP is characterized by its position $\textbf{r}_i(t)$ and orientation $\theta_i(t)$ at any instant $t$. Starting from random orientation, each particle follows Vicsek dynamics, i.e. tries to align with the mean orientation of its neighbours within an interaction radius $R_o$ with some error, i.e. noise. The position and orientation update of $i^{th}$ particle is given by:
\begin{equation}
    \textbf{r}_i(t+1) = \textbf{r}_i(t) + v_o\hat{\textbf{n}}(t)\Delta t
    \label{position}
\end{equation}
\begin{equation}
    \hat{\textbf{n}}_i(t+1) = \frac{\sum_{j\in R_o}\hat{\textbf{n}}_j(t) + \eta k_i(t) \xi_i(t)} {w_i(t)}
    \label{orientation}
\end{equation}
where, $\hat{\textbf{n}}(t) = (cos(\theta_i(t)),sin(\theta_i(t)))$ is the directional unit vector at instant $t$. In Eq.~\ref{orientation}, $\eta$ is the strength of noise (error), $\eta \in [0, 1]$, $k_i(t)$ is the number of neighbours within the interaction radius of the $i^{th}$ particle, and $\xi_i(t)$ is a random unit vector.

\begin{figure} 
\centering
{\includegraphics[width=0.95 \linewidth]{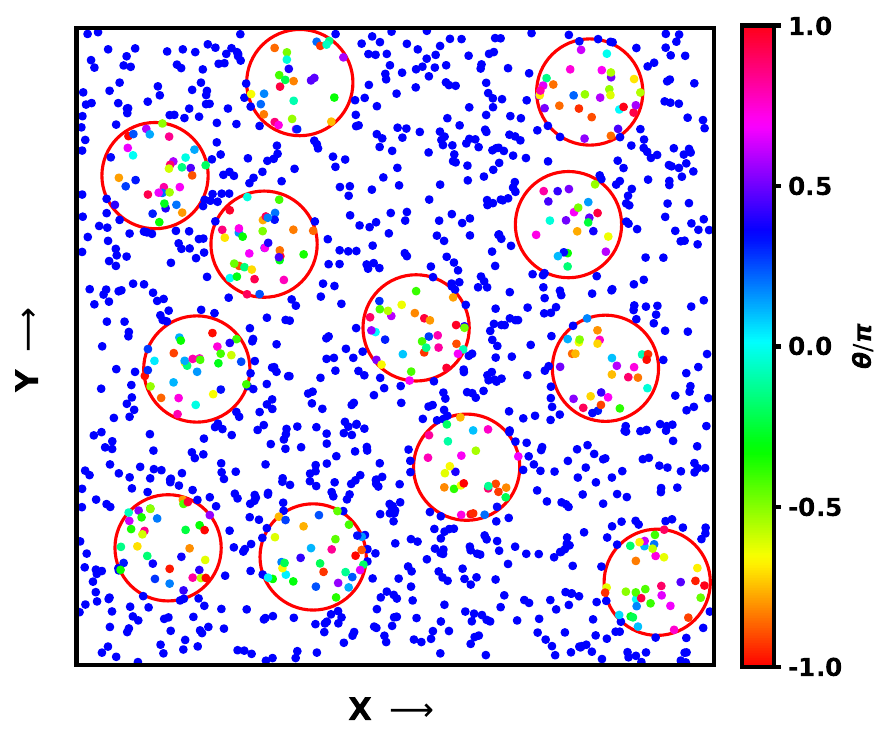}}
    \caption{(color online) Model picture: Self-propelled particles are moving in two-dimensional space. The red circles represent the high noise regimes, while the particle's color represents its orientation, as shown in the color bar. Inside the circles, all particles have varied colors because of high noise, while outside, all SPPs have the same color and exhibit coherent motion.}
\label{Model_picture}
\end{figure}

The spatial inhomogeneity is introduced by $N_o$ non-overlapped circular regions of radius $R_o$ in the space, inside which the strength of noise is maximum (${\eta}_o = 1.0$). Unlike the physical disorder, these regions are accessible to SPPs, but they will impede the flock's ordering due to higher noise strength. In the remaining space, the noise strength is low, which eases the flocking as illustrated in Fig.~\ref{Model_picture}

\textit{Simulation Details}: In the numerical simulation, we consider $R_o=1$, $v_o=0.5$, $\Delta t = 1.0$ and the number density of SPPs $\rho_{n} = N/L^2$ is fixed to $1.0$. We have two varying parameters, the packing fraction of spatial inhomogeneity $\rho_o=N_o\pi R_o^2/L^2$ and the strength of noise $\eta$ in outer space. $\rho_o \in [0,0.75]$ and $\eta \in [0,0.7]$. The simulation starts with a homogeneous distribution of SPPs with random orientations. One simulation step is computed when all the SPPs are updated once according to Eqs.~\ref{position} and ~\ref{orientation}. 
The system evolves over $5\times 10^5$ simulation steps and averages over the last $5 \times 10^4$ steps in the steady state. The Number of particles varies from $N = 2500$ to $90000$, and up to $150$ independent realizations are used for better statistics. 
Averaged normalized velocity $\varphi$ tells the polar ordering in the system. It varies from zero in the disordered phase to unity in the ordered phase.
\begin{equation}
    \varphi(t) = \frac{1}{N}\left|{\sum_{j=1}^{N}{\hat{\textbf{n}}_j(t)} }\right|
    \label{order_parameter}
\end{equation}

\begin{figure} 
\centering
{\includegraphics[width=0.95 \linewidth]{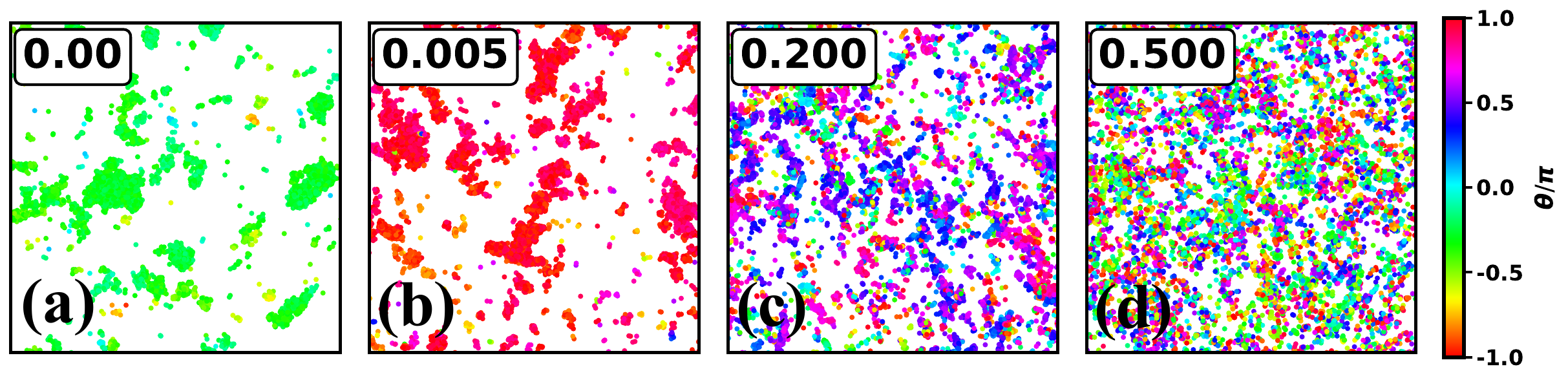}}
    \caption{(color online) Simulation snapshots of the system in the steady-state at $\eta = 0.2$ for four different values of $\rho_o$ as shown in (a),(b),(c) and (d). The orientation of the self-propelled particles (SPPs) is indicated by the color bar. Here $N = 10000$, and for clarity of the figure, the disorders are not shown. \cite{URL}}
\label{system_snapshot}
\end{figure}

\textit{\textbf{Results}}: First, we present the simulation snapshots of the system at four different $\rho_o$ values in Fig.~\ref{system_snapshot}(a-d), which illustrate the effect of inhomogeneity on orientational ordering while keeping the noise fixed at $0.2$. Each circle represents the SPPs, and color of the circle is their orientation $\theta$. For $\rho_o = 0.0$ and $0.005$, the SPPs are moving in one direction in a coherent manner (as shown by the color bar). At $\rho_o = 0.2$, the coherent motion is perturbed, and at $\rho_o = 0.5$, the SPPs are completely mixed and random. The corresponding movies are given in \cite{URL}(MV1-MV4). 

\begin{figure} 
\centering
{\includegraphics[width=0.95 \linewidth]{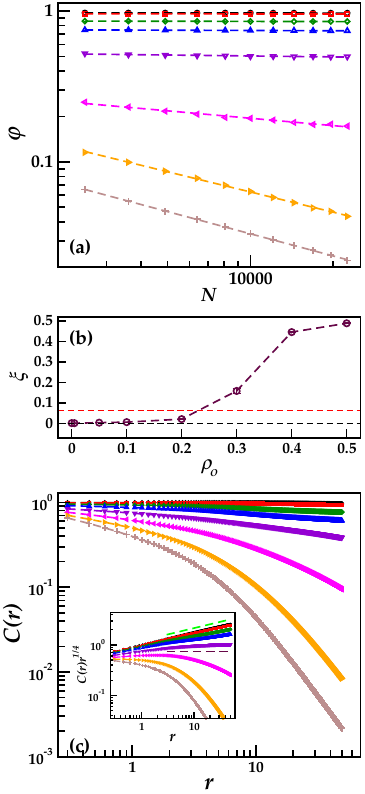}}
    \caption{(color online) (a) Order parameter $\varphi$ vs. number of particle $N$ for various packing fractions $\rho_o$ on log-log scale, for $\eta = 0.2$. 
    The symbol-color scheme: black circle ($\rho_{0} = 0.0$), red square ($\rho_{0} = 0.005$), green diamond ($\rho_{0} = 0.05$), blue upward triangle ($\rho_{0} = 0.1$), violet downward triangle ($\rho_{0} = 0.2$), magenta left-pointing triangle ($\rho_{0} = 0.3$), orange right-pointing triangle ($\rho_{0} = 0.4$), and brown plus sign ($\rho_{0} = 0.5$). The symbols are data points, and dashed lines are respective power-law fits: $\varphi\sim N^{-\zeta(\rho_o)}$. The size of the error bars is less than the size of the symbol. (b) Variation in the exponent $\zeta$ for different $\rho_o$. The red and black dashed lines represent $\zeta=0$ and $1/16$, respectively. (c) Two-point spatial correlation function of the orientation of polar particles $C(r)$ for several packing fractions $\rho_o$ on a log-log scale. \textit{Inset}: y-axis rescaled via multiplication of expected critical exponent value $1/4$. For references, dashed green and maroon lines are drawn with slopes of $0.25$ and $0.0$, respectively. For clarity, the color and symbol scheme used here is the same as that introduced in Fig.~\ref{Phi_vs_N}(a). Here $N=40000$ and $\eta=0.2$}
\label{Phi_vs_N}
\end{figure}

\textbf{Orientational fluctuations and Ordering in the system}: Earlier studies have confirmed that in the clean system of SPPs with alignment interaction, there is evident of true long-range ordered (LRO)  phase in two dimensions \cite{vicsek1995novel,toner1995long}, i.e. for fixed noise, the value of the mean order parameter $\varphi$ does not change with system size $N$. But in the presence of physical disorder, $\varphi$ exhibit power-law decays with $N$: $\varphi \sim N^{-\xi(\rho_o)}$, i.e. existence of quasi-long-range order (QLRO) \cite{chepizhko2013optimal,das2018polar,toner2018hydrodynamic,kosterlitz1973ordering}. In Fig.~\ref{Phi_vs_N}(a), the variation of $\varphi$ with $N$ for multiple $\rho_o$ is shown in log-log scale at fixed noise $\eta=0.2$. For all the cases $\varphi \sim N^{-\xi(\rho_o)}$. In Fig.~\ref{Phi_vs_N}(b), the variation of power $\xi$ is shown. For $\rho_o=0.0$, and $0.005$, $\xi \approx 0$, existence of LRO, and for $\rho_o = 0.05,0.1, 0.2$, $0< \xi < 1/16$ confirms QLRO \cite{chepizhko2013optimal}. At higher $\rho_o$, $\xi$ increases and approaches $1/2$; the system is fully disordered. So as $\rho_o$ varies, the system goes from LRO to disordered via QLRO phase at intermediate $\rho_o$. 

To further confirm the change in the ordering of polar flock due to inhomogeneity, we have calculated the two-point spatial correlation function of the orientation of SPPs $C(r)$ \cite{frenkel1985evidence,chate2006simple,das2017order,mondal2025dynamical}. The correlation is defined as: $C(r)= \langle cos(\theta(r_o+r) - \theta(r_o)) \rangle$, where $\langle..\rangle$ means the average over all the SPPs, different times in the steady state, as well as over multiple realizations. To compute it, we select a particle at position $r_o$ with orientation $\theta(r_o)$, regarded it as the center of the circle, and then computed $cos(\theta(r_o+r) - \theta(r_o))$, where $\theta(r_o + r)$ is the orientation of other particles that reside inside the circle of radius $r$. The calculation is repeated by increasing $r$ up to $L/4$.
In Fig.~\ref{Phi_vs_N}(c), $C(r)$ vs. $r$ is depicted for various $\rho_o$ while keeping noise fixed at $\eta = 0.2$. Similar to Fig.~\ref{Phi_vs_N}(a), for $\rho_o = 0.0$ and $0.005$, $C(r)$ does not show any significant decay as $r$ increases. This tells the consistency of the system with LRO. For $\rho_o = 0.05, 0.1$ and $0.2$, $C(r)$ shows algebraic decays with $r$, which supports the existence of QLRO in the system at a moderate packing fraction of spatial inhomogeneity. As $\rho_o$ further increases, $C(r)$ shows exponential decay, i.e. the system exhibits a disordered phase. $C(r)$ also gives an estimate of the order-disorder transition. Near the transition point, rescaled $C'(r) = C(r)r^{1/4}$ \cite{frenkel1985evidence,goldenfeld2018lectures,chate2006simple} becomes parallel to the x-axis, as shown in Fig.~\ref{Phi_vs_N}(c) $(Inset)$. 


Both results, the finite-size analysis of the order parameter $\varphi$ and the two-point correlation function of the orientation of SPPs $C(r)$ as well as the distribution of orientation for different $\rho_o$ ~\ref{AA} corroborate that the system exhibits a LRO-QLRO-Disorder transition as $\rho_o$ varies from zero to larger values, at a fixed low noise strength $\eta = 0.2$. 

At low $\rho_o$, the existence of LRO is due to the positive feedback of the density on the orientational field. As depicted in Fig.~\ref{system_snapshot}(b), there are dense clusters moving in space with where occasionally they experience regions of $\rho_o$ with high noise. Once they experience such regions, their high density and larger size enable them to overcome the effect of high noise, and only few particles get perturbed, shown in movie \cite{URL}(MV5) (Interaction of a flock with single a circle of high noise). We quantify the variation in the density of particles $\sigma$ inside the high noise regions for different $\rho_o$ and fixed $\eta = 0.2$. We found $\sigma$ shows a non-monotonic trend with respect to $\rho_o$. Both for small and large $\rho_o$ is low and shows a maximum for intermediate $\rho_o$ values, and is shown in  Fig.~\ref{mean_var}. Clearly, the non-monotonicity in $\sigma$ matches well with the three types of ordering: LRO, QLRO, and Disorder. The details of the $\sigma$ is given in Appendix ~\ref{CC}\\
Although the interaction of SPPs with high noise regions differs significantly from physical disorder, the overall effect is remarkably similar to that of physical disorder. For a very low packing fraction, a well-formed flock of SPPs can sustain the perturbation created by the inhomogeneous regions and hence LRO persist in the system. 
Although the nature of ordering is affected due to inhomogeneity, for LRO and QLRO, the number fluctuation is large and reduces slowly as the system approaches the disorder phase, discussed in Appendix~\ref{BB}. Now we analyze the nature of the phase transition from the ordered to the disordered phase by varying the noise strength $\eta$ and $\rho_o$.

\textbf{Nature of Phase Transition}: Equilibrium as well as in nonequilibrium systems, it has been observed that inhomogeneity, whether it comes from interaction among the particles or is present in the substrate, changes the nature of the phase transition \cite{cardy1999quenched,chatelain2001softening,villa2014quenched,singh2021bond,guisandez2017heterogeneity}.

\begin{figure} 
\centering
{\includegraphics[width=0.95\linewidth]{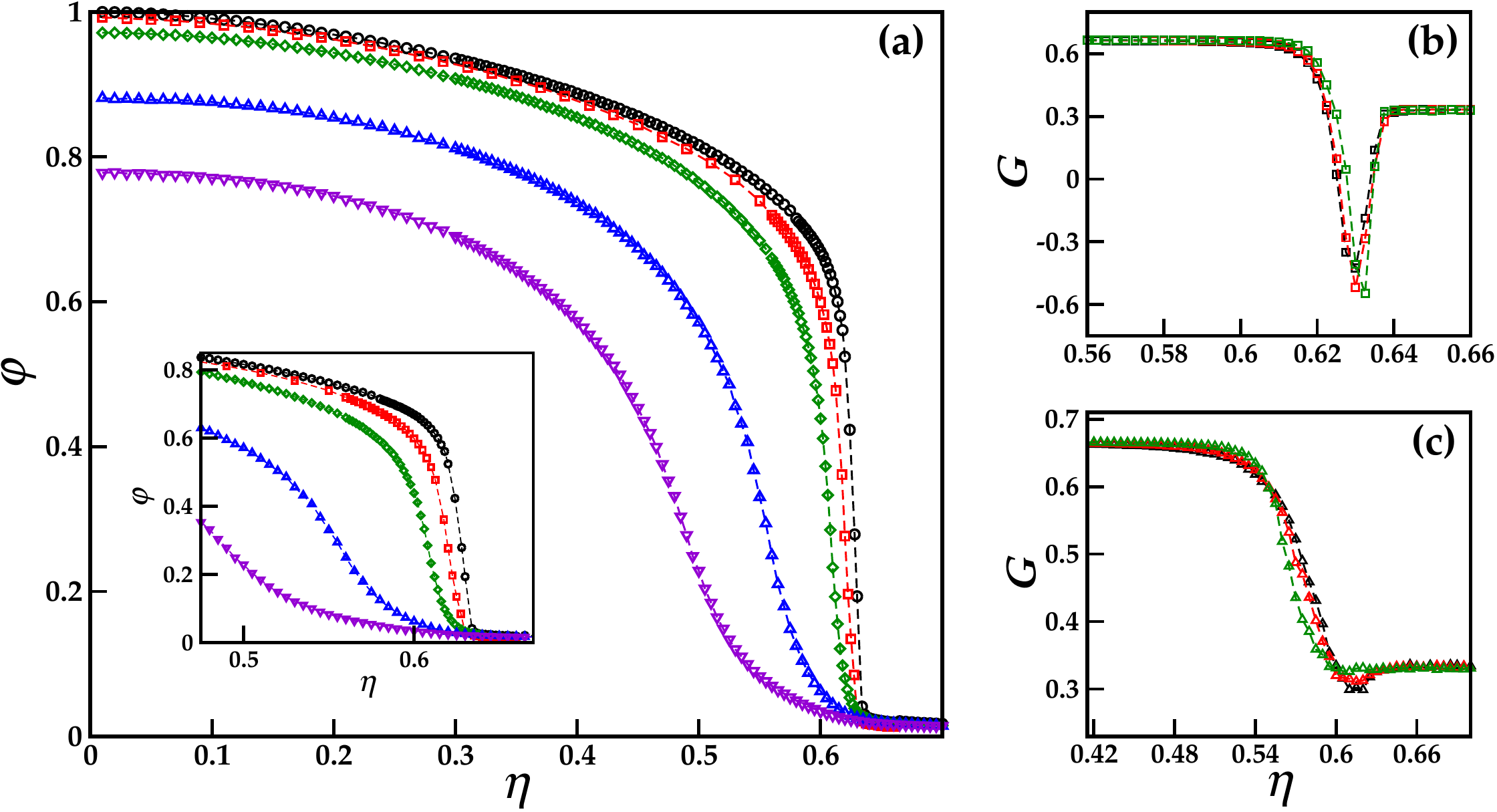}}
    \caption{(color online) (a) Variation of order parameter $\varphi$ with respect to noise $\eta$ for five different values of $\rho_o$. Each curve is color-coded: black circle ($\rho_{0} = 0.0$), red square ($\rho_{0} = 0.0025$), green diamond ($\rho_{0} = 0.01$), blue upward triangle ($\rho_{0} = 0.05$), violet downward triangle ($\rho_{0} = 0.1$). \textit{Inset}: The same plot zoomed near the transition point. Here $N = 10000$. Plots (b) and (c) illustrate system size analysis of Binder cumulant $G$ for two different $\rho_o = 0.0025$ and $0.05$. Three different colors-black, red, and green-represent three system sizes: $N = 2500$, $N = 4900$, and $N = 19600$, respectively. }
\label{Phase_transition}
\end{figure}

From numerous prior numerical studies \cite{gregoire2004onset,chate2008collective,singh2021bond}, it has been well established that a clean system of SPPs, interacting through alignment interaction, exhibits a discontinuous phase transition from a randomly oriented phase with zero ordering to a polar ordered phase with finite ordering as $\eta$ is varied from high to low values. We have displayed order parameter $\varphi$ variation with respect to noise $\eta$ for various $\rho_o$, in Fig.~\ref{Phase_transition}(a). For $\rho_o = 0.0$ and $0.0025$, the variation of $\varphi$ is similar, i.e. discontinuity near the transition point. As $\rho_o$ increases, this variation becomes continuous, and ordering decreases due to the presence of inhomogeneity. So, the polar flock represents deviation from the conventional discontinuous nature of the flocking transition in the presence of inhomogeneity. Also, no optimal noise is found, as reported in previous studies of polar flock in the presence of physical disorder \cite{chepizhko2013optimal,das2018polar}. This subtle difference is mainly due to the intrinsic difference in the interaction of the flock with inhomogeneity.

To further analyze the change in the variation of $\varphi$ in terms of phase transition, we have calculated the fourth-order cumulant of the order parameter, known as the Binder Cumulant $G = 1-\frac{\langle \varphi^4\rangle }{3 {\langle \varphi^2 \rangle}^2}$, a useful quantity to characterize the nature of phase transition in nonequilibrium systems \cite{chate2008collective,kumar2024synchronous}. Here $\langle...\rangle$ signifies the time average. For a 2-D system, it varies from $1/3$ in the disordered phase to $2/3$ in the ordered phase, with a sharp dip towards negative values near the transition point (representing the coexistence of ordered and disordered phases) in the case of a discontinuous one, and a continuous variation in the continuous phase transition. In Fig.~\ref{Phase_transition}(b), Binder cumulant $G$ variation with $\eta$ is shown for very low $\rho_o = 0.0025$, $G$ varies from $1/3$ (disordered) to $2/3$ (ordered) with a sharp negative dip near the transition point ($\varphi$ showing a jump near the transition point as depicted in Fig.~\ref{Phase_transition}(a)), while for a moderate packing fraction $\rho_o = 0.05$, it varies continuously between $1/3$ to $2/3$ ($\varphi$ also varies continuously from disorder to order phase in Fig.~\ref{Phase_transition}(a)) as shown in Fig.~\ref{Phase_transition}(c). The variation of $G$ is consistent across different system sizes depicted in Fig.~\ref{Phase_transition}(b-c). Hence, it is clear that at very low $\rho_o$, the system is still consistent with the discontinuous phase transition, similar to \cite{chepizhko2013optimal}. As $\rho_o$ increases, the system deviates from the discontinuous one and behaves as a continuous transition from QLRO to disorder. 

\begin{figure}
\centering
{\includegraphics[width=0.95 \linewidth]{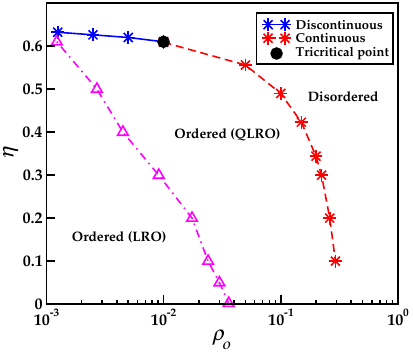}}
    \caption{(colour online) This phase diagram in the $\rho_o-\eta$ plane, with semi-log x scale, represents the ordering of the Vicsek model in the presence of spatial inhomogeneity. The blue solid line with stars shows the discontinuous transition at low $\rho_o$, while the red dashed line with stars represents the continuous phase transition. Both blue solid and red dashed line meets at a black solid filled circle known as a tri-critical point at $\rho_o \approx 0.012$. The ordered phase has two regimes: LRO and QLRO, separated by the dash-dotted magenta line with a triangle up symbol.  }
\label{Phase_diagram}
\end{figure}

Based on these observations, we present a phase diagram for the ordering in the system in the $\rho_o - \eta$ plane as illustrated in Fig.~\ref{Phase_diagram}. We found that the system exhibits LRO in a small regime of the phase diagram at low $\rho_o$. After it, there is a regime of QLRO at moderate $\rho_o$ before reaching disorder at large $\rho_o$. Since the high noise regions introduce additional noise in the system, as $\rho_o$ increases, the order-disorder transition point shifts towards a lower value of $\eta$. Additionally, a larger $\rho_o$ changes the nature of the flocking transition from discontinuous to continuous type.

\textit{\textbf{Conclusion}}: In this work, we have explored a system of polar self-propelled particles on a novel type of inhomogeneous environment. Typically, to create a heterogeneous environment, a portion of the environment is excluded for the particles, thereby introducing physical disorder. But there are many more possibilities to make it inhomogeneous. Singh \textit{et al} \cite{singh2023current} have studied a collection of SPPs moving in a rectangular channel, analogous to the Josephson junction (JJ), where noise inside the junction is very high (to disturb the flocking) and outside the junction it is very low. Due to this geometry, SPPs are reflected from the interface walls of the junction. Motivated by this study, we have investigated a system of SPPs that interact via short-range Vicsek alignment interactions, moving in an inhomogeneous environment. The inhomogeneity is introduced through some randomly placed circular regions inside which the noise is kept fixed at a high value (above critical noise for SPPs), such that SPPs lose cohesion inside them; in the rest of the space, the value of noise is varied from zero to very high. \\

For SPPs, it has also been well established that, for a clean environment, SPPs with Vicsek alignment exhibit LRO and QLRO in the presence of physical disorder in two dimensions. Through a finite-size analysis of the order parameter and two-point correlation of the orientation of SPPs, we also found a transition from LRO to QLRO. Furthermore, there is a change in the nature of the flocking transition from discontinuous to continuous in the presence of inhomogeneity. Upon examining this current study and previous studies of SPPs with physical disorders, we have found that, although both types of inhomogeneity —this one and physical —differ significantly in their interaction with SPPs, the overall effect on the flock is qualitatively similar.

In this study, we have attempted to explore the ordering of polar SPPs in the $\rho_o - \eta$ plane; however, there are still possible extensions that we are seeking. Since we have fixed the size of the spatially inhomogeneous region to be equal to the interaction radius, which is the key parameter determining the length in the system. The variation in size will affect the ordering of the system, introducing a new axis in the phase diagram.     

A.K. thanks Jay Prakash Singh for the helpful discussions. A.K., V.S. and S.M. thank PARAM Shivay for the computational facility under the National Supercomputing Mission, Government of India, at the Indian Institute of Technology (BHU) Varanasi. A.K. thanks PMRF, INDIA, for the research fellowship. S.M. thanks DST, SERB (INDIA), Project No. CRG/2021/006945 and MTR/2021/000438 for partial financial support.\\

The data are available from the authors upon reasonable request.


\appendix
\section{Orientational Distribution}{\label{AA}}

\begin{figure*} 
\centering
{\includegraphics[width=0.95 \linewidth]{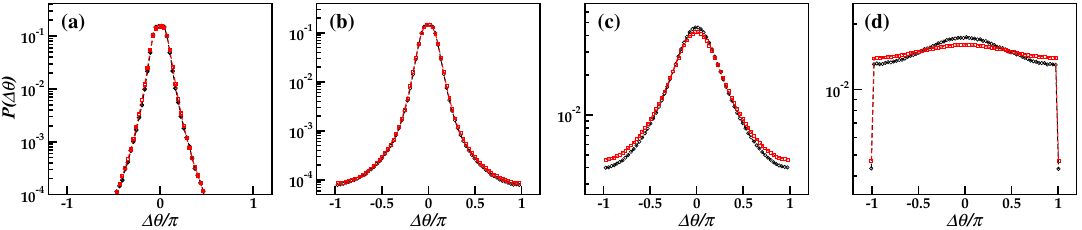}}
    \caption{(colour online) Probability distribution of orientation fluctuation about the mean orientation of flock $P(\Delta\theta)$ is illustrated corresponding to the subplots of Fig.~\ref{system_snapshot} for two different system sizes. The symbol-color scheme: black circle and red squares represent $N=2500, 10000$, respectively.}
\label{Orien_hist}
\end{figure*}

We have also calculated the probability distribution of orientation fluctuation about the mean orientation of the flock $P(\Delta\theta)$, where $\Delta\theta = \theta_j-\overline{\theta}$, here, $\theta_j$ and $\overline{\theta}$ represent the orientation of the $j^{th}$ particle and the mean orientation of the flock, respectively, at any instant $t$ and further the distributions are averaged over many snapshots in the steady state. It also helps to distinguish the change in nature of ordering. In Figs.~\ref {Orien_hist}(a-d), we show $P(\Delta \theta)$ corresponding to Figs.~\ref {system_snapshot}(a-d), and for each case, we compare the results for two different system sizes: $N = 2500$ and $ N = 10000$. \\

For $\rho_o = 0.0$ and $0.005$, we find that the curves for different system sizes $N$ almost overlap. This suggests that there is no significant change in ordering with increasing system size, indicating the presence of long-range order (LRO). However, for $\rho_o = 0.2$, the curve for larger $N$ values broadens, and the peak value shifts to lower values. This suggests that as the system size increases, the ordering decreases, confirming the existence of quasi-long-range order (QLRO) \cite{das2018polar}. At $\rho_o = 0.5$, the distribution approaches uniformity for large $N$, resembling the disordered phase in the system.

\section{Number Fluctuation}
\begin{figure}
\centering
{\includegraphics[width=0.95 \linewidth]{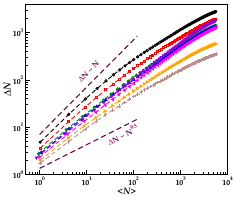}}
    \caption{(color online) Number fluctuation $\Delta N $ vs. the average number of particles $\langle N \rangle$ for multiple obstacle densities $\rho_o$ on log-log scale. The system obeys the relation $\Delta N \sim \langle N \rangle^\alpha$. The color and symbol scheme used here is the same as that introduced in Fig.~\ref{Phi_vs_N}. Here $N=22500$ and $\eta=0.2$. }
\label{Num_fluc}
\end{figure}

{\label{BB}}
Giant number fluctuations (GNF) \cite{ramaswamy2003active, mishra2006active}, a key characteristic of the SPPs system, tell about the number fluctuations in the system. To extract the number fluctuation numerically, we start from the center of the box with a small square of length $l_o = 1$, calculate the number fluctuation $\Delta N = \sqrt{\langle N^2 \rangle - \langle N \rangle^2}$, where $\langle..\rangle$ represents averaged over time and this process is repeated by increasing  $l_o$ up to $L/4$. $\Delta N$ scales as power-law with $\langle N \rangle$: $\Delta N \sim \langle N \rangle^{\alpha}$, where $\alpha = 1/2$ for equilibrium system and $\alpha > 1/2$ for active systems in two dimensions \cite{dey2012spatial, ramaswamy2003active, chate2006simple, mishra2006active}. In Fig.~\ref{Num_fluc}, we have shown $\Delta N$ vs. $\langle N \rangle$ for several $\rho_o$ values for a fixed noise value $\eta = 0.2$. For $\rho = 0.0$, $\alpha \approx 0.88$, i.e. existence of GNF. As $\rho_o$ increases, $\alpha$ continuously decreases, exhibiting suppression of number fluctuation due to inhomogeneity. For all the depicted $\rho_o$, it is always greater than $1/2$ even at very high $\rho_o = 0.5$, where $\varphi \approx 0$, i.e there are small clusters with random orientations as can be seen in \cite{URL}(MV4) and Fig. \ref{system_snapshot}(d) in the system.

\section{Particle fraction inside the inhomogeneous region}{\label{CC}}
\begin{figure}
\centering
{\includegraphics[width=0.95 \linewidth]{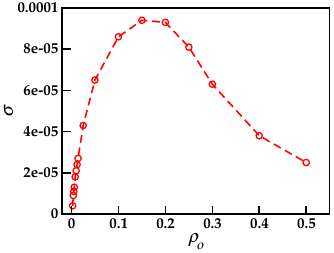}}
    \caption{(color online) Plot shows the variance of the distribution of the number of particles $\sigma$ inside the high noise regions. Here $N=22500$ and $\eta=0.2$. }
\label{mean_var}
\end{figure}

To check how much fraction of SPPs interact with high noise circular regions, we have calculated the distribution of the number of SPPs that are inside these regions at each time step and also averaged it over different time steps in steady state, as well as different realizations. In Fig.~\ref{mean_var}, we have shown the variation in variance of the distribution over a range of $\rho_o$. It shows a non-monotonic variation, which represents that the distribution becomes wider at moderate $\rho_o$. At very low $\rho_o$, there are dense clusters in the system, so whenever it meets with high noise regions, it retains its density as visible from the \cite{URL}(MV2 and MV5). So $\sigma$ is low. As well as the existence of LRO in this $\rho_o$ range, it is also due to very few collisions with the high noise regions and dense clusters, surpassing the effect of high noise due to positive feedback of density. At moderate $\rho_o$, due to a sufficient number of high noise regions, interaction of SPPs with them also becomes frequent, which leads to the dilution of the cluster up to some extent, so the system has a dense cluster (not dense as before) as well as a cluster of a few particles, as shown in Fig.~\ref{system_snapshot}(c). The smaller clusters are again easily perturbed by high noise regions, which lowers the ordering in the system, resulting in a wider distribution in this regime. Here we get QLRO. At high $\rho_o$, due to frequent interactions, SPPs have a homogeneous distribution in the system, so the distribution again shrinks. Since particles are randomized from all over the space, the system reaches the disorder phase.

\section{Description of Movies}
In movies, we have shown the animation of a polar flock moving in a spatially inhomogeneous environment with fixed $N = 10000$ and $ \eta = 0.2$, and varied $\rho_o$.
Parameters for Movies(MV1-MV5) \cite{URL}: \\
Movie1(MV1): $\rho_o = 0.0$\\
Movie2(MV2, MV5): $\rho_o = 0.005$\\
Movie3(MV3): $\rho_o = 0.2$\\
Movie4(MV4): $\rho_o = 0.5$\\

\bibliography{cite1}

\begin{thebibliography}{25}%
\makeatletter
\providecommand \@ifxundefined [1]{%
 \@ifx{#1\undefined}
}%
\providecommand \@ifnum [1]{%
 \ifnum #1\expandafter \@firstoftwo
 \else \expandafter \@secondoftwo
 \fi
}%
\providecommand \@ifx [1]{%
 \ifx #1\expandafter \@firstoftwo
 \else \expandafter \@secondoftwo
 \fi
}%
\providecommand \natexlab [1]{#1}%
\providecommand \enquote  [1]{``#1''}%
\providecommand \bibnamefont  [1]{#1}%
\providecommand \bibfnamefont [1]{#1}%
\providecommand \citenamefont [1]{#1}%
\providecommand \href@noop [0]{\@secondoftwo}%
\providecommand \href [0]{\begingroup \@sanitize@url \@href}%
\providecommand \@href[1]{\@@startlink{#1}\@@href}%
\providecommand \@@href[1]{\endgroup#1\@@endlink}%
\providecommand \@sanitize@url [0]{\catcode `\\12\catcode `\$12\catcode `\&12\catcode `\#12\catcode `\^12\catcode `\_12\catcode `\%12\relax}%
\providecommand \@@startlink[1]{}%
\providecommand \@@endlink[0]{}%
\providecommand \url  [0]{\begingroup\@sanitize@url \@url }%
\providecommand \@url [1]{\endgroup\@href {#1}{\urlprefix }}%
\providecommand \urlprefix  [0]{URL }%
\providecommand \Eprint [0]{\href }%
\providecommand \doibase [0]{http://dx.doi.org/}%
\providecommand \selectlanguage [0]{\@gobble}%
\providecommand \bibinfo  [0]{\@secondoftwo}%
\providecommand \bibfield  [0]{\@secondoftwo}%
\providecommand \translation [1]{[#1]}%
\providecommand \BibitemOpen [0]{}%
\providecommand \bibitemStop [0]{}%
\providecommand \bibitemNoStop [0]{.\EOS\space}%
\providecommand \EOS [0]{\spacefactor3000\relax}%
\providecommand \BibitemShut  [1]{\csname bibitem#1\endcsname}%
\let\auto@bib@innerbib\@empty
\bibitem [{\citenamefont {Vicsek}\ \emph {et~al.}(1995)\citenamefont {Vicsek}, \citenamefont {Czir{\'o}k}, \citenamefont {Ben-Jacob}, \citenamefont {Cohen},\ and\ \citenamefont {Shochet}}]{vicsek1995novel}%
  \BibitemOpen
  \bibfield  {author} {\bibinfo {author} {\bibfnamefont {T.}~\bibnamefont {Vicsek}}, \bibinfo {author} {\bibfnamefont {A.}~\bibnamefont {Czir{\'o}k}}, \bibinfo {author} {\bibfnamefont {E.}~\bibnamefont {Ben-Jacob}}, \bibinfo {author} {\bibfnamefont {I.}~\bibnamefont {Cohen}}, \ and\ \bibinfo {author} {\bibfnamefont {O.}~\bibnamefont {Shochet}},\ }\href@noop {} {\bibfield  {journal} {\bibinfo  {journal} {Physical review letters}\ }\textbf {\bibinfo {volume} {75}},\ \bibinfo {pages} {1226} (\bibinfo {year} {1995})}\BibitemShut {NoStop}%
\bibitem [{\citenamefont {Toner}\ and\ \citenamefont {Tu}(1995)}]{toner1995long}%
  \BibitemOpen
  \bibfield  {author} {\bibinfo {author} {\bibfnamefont {J.}~\bibnamefont {Toner}}\ and\ \bibinfo {author} {\bibfnamefont {Y.}~\bibnamefont {Tu}},\ }\href@noop {} {\bibfield  {journal} {\bibinfo  {journal} {Physical review letters}\ }\textbf {\bibinfo {volume} {75}},\ \bibinfo {pages} {4326} (\bibinfo {year} {1995})}\BibitemShut {NoStop}%
\bibitem [{\citenamefont {Chepizhko}\ \emph {et~al.}(2013)\citenamefont {Chepizhko}, \citenamefont {Altmann},\ and\ \citenamefont {Peruani}}]{chepizhko2013optimal}%
  \BibitemOpen
  \bibfield  {author} {\bibinfo {author} {\bibfnamefont {O.}~\bibnamefont {Chepizhko}}, \bibinfo {author} {\bibfnamefont {E.~G.}\ \bibnamefont {Altmann}}, \ and\ \bibinfo {author} {\bibfnamefont {F.}~\bibnamefont {Peruani}},\ }\href@noop {} {\bibfield  {journal} {\bibinfo  {journal} {Physical review letters}\ }\textbf {\bibinfo {volume} {110}},\ \bibinfo {pages} {238101} (\bibinfo {year} {2013})}\BibitemShut {NoStop}%
\bibitem [{\citenamefont {Das}\ \emph {et~al.}(2018)\citenamefont {Das}, \citenamefont {Kumar},\ and\ \citenamefont {Mishra}}]{das2018polar}%
  \BibitemOpen
  \bibfield  {author} {\bibinfo {author} {\bibfnamefont {R.}~\bibnamefont {Das}}, \bibinfo {author} {\bibfnamefont {M.}~\bibnamefont {Kumar}}, \ and\ \bibinfo {author} {\bibfnamefont {S.}~\bibnamefont {Mishra}},\ }\href@noop {} {\bibfield  {journal} {\bibinfo  {journal} {Physical Review E}\ }\textbf {\bibinfo {volume} {98}},\ \bibinfo {pages} {060602} (\bibinfo {year} {2018})}\BibitemShut {NoStop}%
\bibitem [{\citenamefont {Toner}\ \emph {et~al.}(2018)\citenamefont {Toner}, \citenamefont {Guttenberg},\ and\ \citenamefont {Tu}}]{toner2018hydrodynamic}%
  \BibitemOpen
  \bibfield  {author} {\bibinfo {author} {\bibfnamefont {J.}~\bibnamefont {Toner}}, \bibinfo {author} {\bibfnamefont {N.}~\bibnamefont {Guttenberg}}, \ and\ \bibinfo {author} {\bibfnamefont {Y.}~\bibnamefont {Tu}},\ }\href@noop {} {\bibfield  {journal} {\bibinfo  {journal} {Physical Review E}\ }\textbf {\bibinfo {volume} {98}},\ \bibinfo {pages} {062604} (\bibinfo {year} {2018})}\BibitemShut {NoStop}%
\bibitem [{\citenamefont {Mishra}\ \emph {et~al.}(2023)\citenamefont {Mishra}, \citenamefont {Krishna},\ and\ \citenamefont {Mishra}}]{mishra2023active}%
  \BibitemOpen
  \bibfield  {author} {\bibinfo {author} {\bibfnamefont {P.~K.}\ \bibnamefont {Mishra}}, \bibinfo {author} {\bibfnamefont {A.}~\bibnamefont {Krishna}}, \ and\ \bibinfo {author} {\bibfnamefont {S.}~\bibnamefont {Mishra}},\ }\href@noop {} {\bibfield  {journal} {\bibinfo  {journal} {Soft Materials}\ }\textbf {\bibinfo {volume} {21}},\ \bibinfo {pages} {377} (\bibinfo {year} {2023})}\BibitemShut {NoStop}%
\bibitem [{URL(2024)}]{URL}%
  \BibitemOpen
  \href@noop {} {\bibfield  {journal} {\bibinfo  {journal} {Supplementary Material \url{https://drive.google.com/drive/folders/1sVzyIof7QDQF9_gTau-17-7Wu_PfH1Mm?usp=drive_link}}\ } (\bibinfo {year} {2024})}\BibitemShut {NoStop}%
\bibitem [{\citenamefont {Kosterlitz}\ and\ \citenamefont {Thouless}(1973)}]{kosterlitz1973ordering}%
  \BibitemOpen
  \bibfield  {author} {\bibinfo {author} {\bibfnamefont {J.~M.}\ \bibnamefont {Kosterlitz}}\ and\ \bibinfo {author} {\bibfnamefont {D.~J.}\ \bibnamefont {Thouless}},\ }\href@noop {} {\bibfield  {journal} {\bibinfo  {journal} {Journal of Physics C: Solid State Physics}\ }\textbf {\bibinfo {volume} {6}},\ \bibinfo {pages} {1181} (\bibinfo {year} {1973})}\BibitemShut {NoStop}%
\bibitem [{\citenamefont {Frenkel}\ and\ \citenamefont {Eppenga}(1985)}]{frenkel1985evidence}%
  \BibitemOpen
  \bibfield  {author} {\bibinfo {author} {\bibfnamefont {D.}~\bibnamefont {Frenkel}}\ and\ \bibinfo {author} {\bibfnamefont {R.}~\bibnamefont {Eppenga}},\ }\href@noop {} {\bibfield  {journal} {\bibinfo  {journal} {Physical Review A}\ }\textbf {\bibinfo {volume} {31}},\ \bibinfo {pages} {1776} (\bibinfo {year} {1985})}\BibitemShut {NoStop}%
\bibitem [{\citenamefont {Chat{\'e}}\ \emph {et~al.}(2006)\citenamefont {Chat{\'e}}, \citenamefont {Ginelli},\ and\ \citenamefont {Montagne}}]{chate2006simple}%
  \BibitemOpen
  \bibfield  {author} {\bibinfo {author} {\bibfnamefont {H.}~\bibnamefont {Chat{\'e}}}, \bibinfo {author} {\bibfnamefont {F.}~\bibnamefont {Ginelli}}, \ and\ \bibinfo {author} {\bibfnamefont {R.}~\bibnamefont {Montagne}},\ }\href@noop {} {\bibfield  {journal} {\bibinfo  {journal} {Physical review letters}\ }\textbf {\bibinfo {volume} {96}},\ \bibinfo {pages} {180602} (\bibinfo {year} {2006})}\BibitemShut {NoStop}%
\bibitem [{\citenamefont {Das}\ \emph {et~al.}(2017)\citenamefont {Das}, \citenamefont {Kumar},\ and\ \citenamefont {Mishra}}]{das2017order}%
  \BibitemOpen
  \bibfield  {author} {\bibinfo {author} {\bibfnamefont {R.}~\bibnamefont {Das}}, \bibinfo {author} {\bibfnamefont {M.}~\bibnamefont {Kumar}}, \ and\ \bibinfo {author} {\bibfnamefont {S.}~\bibnamefont {Mishra}},\ }\href@noop {} {\bibfield  {journal} {\bibinfo  {journal} {Scientific Reports}\ }\textbf {\bibinfo {volume} {7}},\ \bibinfo {pages} {7080} (\bibinfo {year} {2017})}\BibitemShut {NoStop}%
\bibitem [{\citenamefont {Mondal}\ \emph {et~al.}(2025)\citenamefont {Mondal}, \citenamefont {Mishra}, \citenamefont {Vicsek},\ and\ \citenamefont {Mishra}}]{mondal2025dynamical}%
  \BibitemOpen
  \bibfield  {author} {\bibinfo {author} {\bibfnamefont {P.~S.}\ \bibnamefont {Mondal}}, \bibinfo {author} {\bibfnamefont {P.~K.}\ \bibnamefont {Mishra}}, \bibinfo {author} {\bibfnamefont {T.}~\bibnamefont {Vicsek}}, \ and\ \bibinfo {author} {\bibfnamefont {S.}~\bibnamefont {Mishra}},\ }\href@noop {} {\bibfield  {journal} {\bibinfo  {journal} {Physica A: Statistical Mechanics and its Applications}\ }\textbf {\bibinfo {volume} {659}},\ \bibinfo {pages} {130338} (\bibinfo {year} {2025})}\BibitemShut {NoStop}%
\bibitem [{\citenamefont {Goldenfeld}(2018)}]{goldenfeld2018lectures}%
  \BibitemOpen
  \bibfield  {author} {\bibinfo {author} {\bibfnamefont {N.}~\bibnamefont {Goldenfeld}},\ }\href@noop {} {\emph {\bibinfo {title} {Lectures on phase transitions and the renormalization group}}}\ (\bibinfo  {publisher} {CRC Press},\ \bibinfo {year} {2018})\BibitemShut {NoStop}%
\bibitem [{\citenamefont {Cardy}(1999)}]{cardy1999quenched}%
  \BibitemOpen
  \bibfield  {author} {\bibinfo {author} {\bibfnamefont {J.}~\bibnamefont {Cardy}},\ }\href@noop {} {\bibfield  {journal} {\bibinfo  {journal} {Physica A: Statistical Mechanics and its Applications}\ }\textbf {\bibinfo {volume} {263}},\ \bibinfo {pages} {215} (\bibinfo {year} {1999})}\BibitemShut {NoStop}%
\bibitem [{\citenamefont {Chatelain}\ \emph {et~al.}(2001)\citenamefont {Chatelain}, \citenamefont {Berche}, \citenamefont {Janke},\ and\ \citenamefont {Berche}}]{chatelain2001softening}%
  \BibitemOpen
  \bibfield  {author} {\bibinfo {author} {\bibfnamefont {C.}~\bibnamefont {Chatelain}}, \bibinfo {author} {\bibfnamefont {B.}~\bibnamefont {Berche}}, \bibinfo {author} {\bibfnamefont {W.}~\bibnamefont {Janke}}, \ and\ \bibinfo {author} {\bibfnamefont {P.~E.}\ \bibnamefont {Berche}},\ }\href@noop {} {\bibfield  {journal} {\bibinfo  {journal} {Physical Review E}\ }\textbf {\bibinfo {volume} {64}},\ \bibinfo {pages} {036120} (\bibinfo {year} {2001})}\BibitemShut {NoStop}%
\bibitem [{\citenamefont {Villa~Mart{\'\i}n}\ \emph {et~al.}(2014)\citenamefont {Villa~Mart{\'\i}n}, \citenamefont {Bonachela},\ and\ \citenamefont {Munoz}}]{villa2014quenched}%
  \BibitemOpen
  \bibfield  {author} {\bibinfo {author} {\bibfnamefont {P.}~\bibnamefont {Villa~Mart{\'\i}n}}, \bibinfo {author} {\bibfnamefont {J.~A.}\ \bibnamefont {Bonachela}}, \ and\ \bibinfo {author} {\bibfnamefont {M.~A.}\ \bibnamefont {Munoz}},\ }\href@noop {} {\bibfield  {journal} {\bibinfo  {journal} {Physical Review E}\ }\textbf {\bibinfo {volume} {89}},\ \bibinfo {pages} {012145} (\bibinfo {year} {2014})}\BibitemShut {NoStop}%
\bibitem [{\citenamefont {Singh}\ \emph {et~al.}(2021)\citenamefont {Singh}, \citenamefont {Kumar},\ and\ \citenamefont {Mishra}}]{singh2021bond}%
  \BibitemOpen
  \bibfield  {author} {\bibinfo {author} {\bibfnamefont {J.~P.}\ \bibnamefont {Singh}}, \bibinfo {author} {\bibfnamefont {S.}~\bibnamefont {Kumar}}, \ and\ \bibinfo {author} {\bibfnamefont {S.}~\bibnamefont {Mishra}},\ }\href@noop {} {\bibfield  {journal} {\bibinfo  {journal} {Journal of Statistical Mechanics: Theory and Experiment}\ }\textbf {\bibinfo {volume} {2021}},\ \bibinfo {pages} {083217} (\bibinfo {year} {2021})}\BibitemShut {NoStop}%
\bibitem [{\citenamefont {Guisandez}\ \emph {et~al.}(2017)\citenamefont {Guisandez}, \citenamefont {Baglietto},\ and\ \citenamefont {Rozenfeld}}]{guisandez2017heterogeneity}%
  \BibitemOpen
  \bibfield  {author} {\bibinfo {author} {\bibfnamefont {L.}~\bibnamefont {Guisandez}}, \bibinfo {author} {\bibfnamefont {G.}~\bibnamefont {Baglietto}}, \ and\ \bibinfo {author} {\bibfnamefont {A.}~\bibnamefont {Rozenfeld}},\ }\href@noop {} {\bibfield  {journal} {\bibinfo  {journal} {arXiv preprint arXiv:1711.11531}\ } (\bibinfo {year} {2017})}\BibitemShut {NoStop}%
\bibitem [{\citenamefont {Gr{\'e}goire}\ and\ \citenamefont {Chat{\'e}}(2004)}]{gregoire2004onset}%
  \BibitemOpen
  \bibfield  {author} {\bibinfo {author} {\bibfnamefont {G.}~\bibnamefont {Gr{\'e}goire}}\ and\ \bibinfo {author} {\bibfnamefont {H.}~\bibnamefont {Chat{\'e}}},\ }\href@noop {} {\bibfield  {journal} {\bibinfo  {journal} {Physical review letters}\ }\textbf {\bibinfo {volume} {92}},\ \bibinfo {pages} {025702} (\bibinfo {year} {2004})}\BibitemShut {NoStop}%
\bibitem [{\citenamefont {Chat{\'e}}\ \emph {et~al.}(2008)\citenamefont {Chat{\'e}}, \citenamefont {Ginelli}, \citenamefont {Gr{\'e}goire},\ and\ \citenamefont {Raynaud}}]{chate2008collective}%
  \BibitemOpen
  \bibfield  {author} {\bibinfo {author} {\bibfnamefont {H.}~\bibnamefont {Chat{\'e}}}, \bibinfo {author} {\bibfnamefont {F.}~\bibnamefont {Ginelli}}, \bibinfo {author} {\bibfnamefont {G.}~\bibnamefont {Gr{\'e}goire}}, \ and\ \bibinfo {author} {\bibfnamefont {F.}~\bibnamefont {Raynaud}},\ }\href@noop {} {\bibfield  {journal} {\bibinfo  {journal} {Physical Review E—Statistical, Nonlinear, and Soft Matter Physics}\ }\textbf {\bibinfo {volume} {77}},\ \bibinfo {pages} {046113} (\bibinfo {year} {2008})}\BibitemShut {NoStop}%
\bibitem [{\citenamefont {Kumar}\ \emph {et~al.}(2024)\citenamefont {Kumar}, \citenamefont {Pattanayak}, \citenamefont {Singh},\ and\ \citenamefont {Mishra}}]{kumar2024synchronous}%
  \BibitemOpen
  \bibfield  {author} {\bibinfo {author} {\bibfnamefont {A.}~\bibnamefont {Kumar}}, \bibinfo {author} {\bibfnamefont {S.}~\bibnamefont {Pattanayak}}, \bibinfo {author} {\bibfnamefont {R.}~\bibnamefont {Singh}}, \ and\ \bibinfo {author} {\bibfnamefont {S.}~\bibnamefont {Mishra}},\ }\href@noop {} {\bibfield  {journal} {\bibinfo  {journal} {Physics Letters A}\ }\textbf {\bibinfo {volume} {523}},\ \bibinfo {pages} {129773} (\bibinfo {year} {2024})}\BibitemShut {NoStop}%
\bibitem [{\citenamefont {Singh}\ \emph {et~al.}(2023)\citenamefont {Singh}, \citenamefont {Mondal}, \citenamefont {Semwal},\ and\ \citenamefont {Mishra}}]{singh2023current}%
  \BibitemOpen
  \bibfield  {author} {\bibinfo {author} {\bibfnamefont {J.~P.}\ \bibnamefont {Singh}}, \bibinfo {author} {\bibfnamefont {P.~S.}\ \bibnamefont {Mondal}}, \bibinfo {author} {\bibfnamefont {V.}~\bibnamefont {Semwal}}, \ and\ \bibinfo {author} {\bibfnamefont {S.}~\bibnamefont {Mishra}},\ }\href@noop {} {\bibfield  {journal} {\bibinfo  {journal} {Physical Review E}\ }\textbf {\bibinfo {volume} {108}},\ \bibinfo {pages} {034608} (\bibinfo {year} {2023})}\BibitemShut {NoStop}%
\bibitem [{\citenamefont {Ramaswamy}\ \emph {et~al.}(2003)\citenamefont {Ramaswamy}, \citenamefont {Simha},\ and\ \citenamefont {Toner}}]{ramaswamy2003active}%
  \BibitemOpen
  \bibfield  {author} {\bibinfo {author} {\bibfnamefont {S.}~\bibnamefont {Ramaswamy}}, \bibinfo {author} {\bibfnamefont {R.~A.}\ \bibnamefont {Simha}}, \ and\ \bibinfo {author} {\bibfnamefont {J.}~\bibnamefont {Toner}},\ }\href@noop {} {\bibfield  {journal} {\bibinfo  {journal} {Europhysics Letters}\ }\textbf {\bibinfo {volume} {62}},\ \bibinfo {pages} {196} (\bibinfo {year} {2003})}\BibitemShut {NoStop}%
\bibitem [{\citenamefont {Mishra}\ and\ \citenamefont {Ramaswamy}(2006)}]{mishra2006active}%
  \BibitemOpen
  \bibfield  {author} {\bibinfo {author} {\bibfnamefont {S.}~\bibnamefont {Mishra}}\ and\ \bibinfo {author} {\bibfnamefont {S.}~\bibnamefont {Ramaswamy}},\ }\href@noop {} {\bibfield  {journal} {\bibinfo  {journal} {Physical review letters}\ }\textbf {\bibinfo {volume} {97}},\ \bibinfo {pages} {090602} (\bibinfo {year} {2006})}\BibitemShut {NoStop}%
\bibitem [{\citenamefont {Dey}\ \emph {et~al.}(2012)\citenamefont {Dey}, \citenamefont {Das},\ and\ \citenamefont {Rajesh}}]{dey2012spatial}%
  \BibitemOpen
  \bibfield  {author} {\bibinfo {author} {\bibfnamefont {S.}~\bibnamefont {Dey}}, \bibinfo {author} {\bibfnamefont {D.}~\bibnamefont {Das}}, \ and\ \bibinfo {author} {\bibfnamefont {R.}~\bibnamefont {Rajesh}},\ }\href@noop {} {\bibfield  {journal} {\bibinfo  {journal} {Physical review letters}\ }\textbf {\bibinfo {volume} {108}},\ \bibinfo {pages} {238001} (\bibinfo {year} {2012})}\BibitemShut {NoStop}%
\end{thebibliography}%

\end{document}